# Zipf's law, Hierarchical Structure, and Shuffling-Cards Model for Urban Development

Yanguang Chen

(Department of Geography, College of Urban and Environmental Sciences, Peking University, Beijing 100871, P.R. China)

**Abstract**: A new angle of view is proposed to find the simple rules dominating complex systems and regular patterns behind random phenomena such as cities. Hierarchy of cities reflects the ubiquitous structure frequently observed in the natural world and social institutions. Where there is a hierarchy with cascade structure, there is a rank-size distribution following Zipf's law, and *vice versa*. The hierarchical structure can be described with a set of exponential functions that are identical in form to Horton-Strahler's laws on rivers and Gutenberg-Richter's laws on earthquake energy. From the exponential models, we can derive four power laws such as Zipf's law indicative of fractals and scaling symmetry. Research on the hierarchy is revealing for us to understand how complex systems are self-organized. A card-shuffling model is built to interpret the relation between Zipf's law and hierarchy of cities. This model can be expanded to explain the general empirical power-law distributions across the individual physical and social sciences, which are hard to be comprehended within the specific scientific domains.
**Key words**: hierarchy of cities; fractal; allometric scaling; Zipf's law; rank-size rule; $2^n$ rule; river network; earthquake energy distribution

## 1. Introduction

The well-known Zipf's law is a very basic principle for city-size distributions, and empirically, the Zipf distribution is always associated with hierarchical structure of urban systems (Chen, 2008). Hierarchy is frequently observed within the natural world as well as in social institutions, and it is a form of organization of complex systems which depend on or produce a strong differentiation in power and size between the parts of the whole (Pumain, 2006). A system of cities in a region is always organized as a hierarchy with cascade structure (Jiang and Yao, 2010). Where mathematical models is concerned, a hierarchy of cities always bears an analogy to network of rivers (Krugman, 1996; Woldenberg and Berry, 1967), while the latter has an analogy with earthquake energy distribution (Chen and Zhou, 2008). There seems to be hidden order behind random distributions of cities, and the similar order can be found behind river networks and earthquake phenomena. Studies on urban hierarchies will be helpful for us to understand the general natural laws which dominate both physical and human systems.

Urban evolution takes on two prominent properties: one is the Zipf distribution at the large scale (Batty, 2006; Batty and Longley, 1994; Chen and Zhou, 2003; Gabaix, 1999), the other is the hierarchical scaling relations at different scales and measures (e.g. Batty, 2008; Carvalho and Penn, 2004; Jiang, 2007; Jiang and Yao, 2010; Isalgue *et al*, 2007; Kuhnert *et al*, 2006; Lammer *et al*, 2006). If a study area is large enough, the size distribution of cities in the area always follows Zipf's law. The Zipf distribution, i.e., the rank-size distribution, is one of ubiquitous general empirical observations across the individual sciences (e.g Adamic and Huberman, 2002; Axtell, 2001; Gabaix *et al*, 2003; Okuyama *et al*, 1999), which cannot be understood with the set of references developed within the specific scientific domain (Bak, 1996). In fact, the Zipf distribution and hierarchical structure is two different sides of the same coin. Hierarchy can provide a new angle of view to understand Zipf's law and allometric scaling of cities, and *vice versa*. Both Zipf's law and allomtric growth law are related with fractals (e.g. Batty and Longley, 1994; Chen and Zhou, 2003; Frankhauser, 1990; Mandelbrot, 1983; Nicolis *et al*, 1989; West, 1999), and fractal theory is one of powerful tools for researching complexity and regularity of urban development.

In this paper, Zipf's law, allometric scaling, and fractal relations will be integrated into the same framework based on hierarchy of cities, and then, a model of playing cards will be proposed to explain the Zipf distribution and hierarchical scaling. From this framework, we can gain an insight into cities in the new perspective. Especially, this theoretical framework and model can be generalized to physical scientific fields. The rest of this paper is organized as follows. In section 2, three exponential models associated with four power laws on hierarchy of cities are presented, and an analogy between cities, rivers, and earthquake energy is drawn to show the ubiquity of hierarchical structure. In section 3, two case analyses based on large-scale urban systems are made to lend further support to power laws and exponential laws of cities. In section 4, a theory of shuffling cards on urban evolution is illustrated to interpret the spatial patterns and hidden rules of city distributions. Finally, the discussion is concluded with several simple comments.

## 2. Cities, rivers, and earthquakes: analogous systems?

### 2.1 The scaling laws of cities

First of all, the mathematical description of hierarchies of cities should be presented here. Grouping the cities in a large-scale region into $M$ classes in a top-down order, we can define an urban hierarchy with cascade structure. The hierarchy of cities can be modeled with a set of exponential equations

$$N_m = N_1 r_n^{m-1}, \tag{1}$$

$$P_m = P_1 r_p^{1-m}, \tag{2}$$

$$A_m = A_1 r_a^{1-m}, \tag{3}$$

where $m$ denotes the top-down ordinal number of city class ($m$=1, 2, $\cdots$, $M$), $N_m$ refers to the number of cities of a given size, correspondingly, $P_m$ and $A_m$ to the mean population size and urban area in the $m$th class. As for the parameters, $N_1$ is the number of the top-order cities, $P_1$ and $A_1$ are the mean population and urban area of the first-order cities. In theory, we take $N_1$=1. The common ratios are defined as follows: $r_n = N_{m+1}/N_m$ denotes the interclass **number ratio** of cities, $r_p = P_m/P_{m+1}$ the population **size ratio**, and $r_a = A_m/A_{m+1}$ the urban **area ratio**. In fact, equations (1) and (2) are just the generalized Beckmann-Davis models (Beckmann, 1958; Davis, 1978; Chen and Zhou, 2003). According to Davis (1978), if $r_n$=2 as given, then it will follow that $r_p \rightarrow 2$, where the arrow denotes "approach" or "be close to". If so, equations (1) and (2) express the $2^n$ rule, otherwise they express the generalized $2^n$ rule.

Several power-law relations can be derived from the above exponential laws. Rearranging equation (2) yields $r_p^{m-1}=P_1/P_m$, then taking logarithm to the base $r_n$ of this equation and substituting the result into equation (1) yields a power function as

$$N_m = \mu P_m^{-D}, \tag{4}$$

where $\mu = N_1 P_1^D$, $D=\ln r_n/\ln r_p$. Equation (4) can be termed as the *size-number scaling relation* of cities, and *D* is just the fractal dimension of urban hierarchies measured with population (Jiang and Yao, 2010). By analogy, the *area-number scaling relation* of cities can be derived from equations (1) and (3) in the form

$$N_m = \eta A_m^{-d}, \tag{5}$$

in which $\eta=N_1 A_1^d$, $d=\ln r_n/\ln r_a$. Here *d* can be regarded as the fractal dimension of urban hierarchies measured with urban area. It is easy for us to derive an allometric scaling relation between urban area and population from equation (2) and (3) such as

$$A_m = a P_m^b, \tag{6}$$

where $a=A_1 P_1^{-b}$ denotes the proportionality coefficient, and $b=\ln r_a/\ln r_p=D/d$ is the scaling exponent. In light of the dimensional consistency, the allometric scaling exponent is actually the ratio of the fractal dimension of urban form to that of urban population.

In theory, the size-number scaling relation, equation (4), is equivalent mathematically to the three-parameter Zipf-type model on size distribution (Chen and Zhou, 2003; Gell-Mann, 1994; Mandelbrot, 1983). The latter can also be derived from equations (1) and (2), and the result is

$$P(\rho) = C(\rho - \varsigma)^{-d_z}, \tag{7}$$

where $\rho$ is the rank of cities in decreasing order of size, $P(\rho)$ is the population of the $\rho$th city. As the parameters, we have the constant of proportionality $C=P_1[r_n/(r_n-1)]^{1/D}$, small parameter $\varsigma=1/(1-r_n)$, and the power exponent $d_z=1/D=\ln r_p/\ln r_n$ (Chen and Zhou, 2003). If we omit the small parameter from equation (7), we have the common two-parameter Zipf model

$$P(\rho) = P_1 \rho^{-q}, \tag{8}$$

where $P_1$ is the population size of the largest city, and *q* the Zipf exponent ($q \approx d_z$). When $q=1$ as given, then we will have the one-parameter Zipf model

$$P(\rho) = \frac{P_1}{\rho}. \tag{9}$$

which is the well-known *rank-size rule* equivalent to the $2^n$ rule on cities. The rank-size distribution suggests self-similarity behind random patterns, and fractal dimension is an important parameter to understand urban hierarchy (Chen and Zhou, 2003; Haag, 1994; Mandelbrot, 1983).

## 2.2 Analogy of cities with rivers and earthquake

The hierarchy of cities reflects the cascade structure which is ubiquitous in both physical and human systems. To provide a general pattern for us to understand how the evolutive systems are self-organized, let's draw an analogy between cities, rivers, and earthquake energy distributions (Figure 1). In fact, equations (1), (2), and (3), have the property of 'mirror symmetry'. That is, if we transpose the order $m$, the structure of mathematical models will not vary, but exponents will change sign. Thus the three exponential laws can be rewritten as follows

$$N_m = N_1 r_n^{1-m}, \tag{10}$$

$$P_m = P_1 r_p^{m-1}, \tag{11}$$

$$A_m = A_1 r_a^{m-1}, \tag{12}$$

where $m$ denotes the bottom-up ordinal number ($m=1, 2, \cdots, M$), $N_m$, $P_m$, and $A_m$ fulfill the same roles as in equations (1), (2), and (3), $N_1$, $P_1$, and $A_1$ represent the city number, population size, and urban area of the bottom order, respectively, and $N_1 \gg 1$ now. As regards the ratio parameter, we have $r_n = N_m/N_{m+1}$, $r_p = P_{m+1}/P_m$, and $r_a = A_{m+1}/A_m$ (Chen and Zhou, 2008).

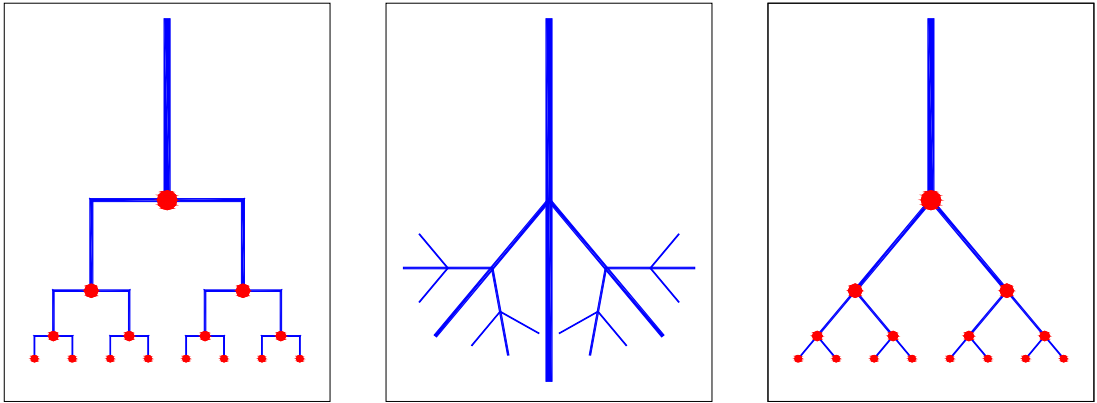

a. Hierarchy of cities  b. Networks of rivers  c. Hierarchy of earthquakes

**Figure 1 The models of hierarchies of cities, rivers, and earthquakes with cascade structure**

(**Note**: The sketch maps only show the first four classes for the top-down models, or the last four classes for the bottom-up models.)

These exponential models can be employed to characterize river networks and hierarchies of the seismic activities of a region (say, Japan) over a period of time (say, 30 years). Equations (10), (11), and (12) bear an analogy to Horton-Strahler's laws in geomorphology (Horton, 1945; Schumm, 1956; Strahler, 1952) and Gutenberg-Richter's laws in geology and seismology (Gutenberg and Richter, 1954; Turcotte, 1997). If the three exponential laws on cities, Horton-Strahler's laws on rivers, and Gutenberg-Richter's laws on earthquake are tabulated for comparison, they are identical in form to one another (Table 1). According to Horton (1945), Schumm (1956), and Strahler (1952), the scaling relations of a network of rivers can measured with river branch length ($L$), the number of tributary rivers of a given length ($B$), and drainage areas ($S$). According to Gutenberg and Richter (1954), a hierarchy of seismic activities can also be described with three measurements: the size of released energy ($E$), the frequency/number of earthquakes of a certain magnitude ($f$), and rupture area ($U_m$). The ordinal number indicative of the class of cities or rivers corresponds to the moment magnitude scale (MMS) of earthquakes. Thus, the similarity between equations (10), (11), and (12) with Horton-Strahler's laws as well as Gutenberg-Richter's laws is based on the corresponding measurement relations as follows: (1) city number ($N_m$) → river branch number ($B_m$) → earthquake frequency ($f_m$); (2) city population size ($P_m$) → river branch/segment length ($L_m$) → earthquake energy ($E_m$); (3) urbanized area ($A_m$) → drainage/catchment area ($S_m$) → fault break area ($U_m$).

**Table 1 Comparison between the exponential laws of cities and those of rivers and earthquake energy**

| Exponential law | Hierarch of cities | Network of rivers | Energy of earthquake |
|---|---|---|---|
| The first law | $N_m = N_1 r_n^{1-m}$ | $B_m = B_1 r_b^{1-m}$ | $f_m = f_1 r_f^{1-m}$ |
| The second law | $P_m = P_1 r_p^{m-1}$ | $L_m = L_1 r_l^{m-1}$ | $E_m = E_1 r_e^{m-1}$ |
| The third law | $A_m = A_1 r_a^{m-1}$ | $S_m = S_1 r_s^{m-1}$ | $U_m = U_1 r_s^{m-1}$ |

**Note**: These exponential laws correspond to the visual models displayed in Figure 1. In Horton-Strahler's law, the ratios are defined as $r_b=B_m/B_{m+1}$, $r_l=L_{m+1}/L_m$, and $r_s=S_{m+1}/S_m$; in Gutenberg-Richter's laws, the ratios are given by $r_f=f_m/f_{m+1}$, $r_e=E_{m+1}/E_m$, and $r_u=U_{m+1}/U_m$.

Despite all these similarities, there are clear differences among cities, rivers, and earthquake energy distributions as hierarchies. Actually, hierarchies can be divided into two types: one is the *real hierarchy* with physical cascade structure such as a system of rivers, and the other is *dummy* *hierarchy* with mathematical cascade structure such as earthquake energy in given period and region. For river systems, the rivers of order *m* have direct connection with those of order ($m\pm1$). However, for earthquake, the quake energy sizes in the *m*th class have no fixed relation to those in the ($m\pm1$)th class. For example, if the MMS of a main shock in a place is 7, the MMS of its foreshocks and aftershocks are usually 3~5 rather than 6. The earthquakes of order 6 and 8 often occur in another place and time and cannot be directly related to the shock of order 7. Generally speaking, the interclass relation in a dummy hierarchy is in the mathematical sense rather than physical sense. Cities come between rivers and earthquakes. It is hard for us to bring to light the physical cascade structure of a hierarchy of cities, but it is convenient to research into its mathematical structure.

Typically, Horton-Strahler's laws are on real hierarchies, while Gutenberg-Richter's laws on dummy hierarchies (Table 2). There are many empirical analyses about Horton-Straler's law and Gutenberg-Richter's laws (Rodriguez-Iturbe and Rinaldo, 2001; Turcotte, 1997). As for the exponential laws of cities, preliminary empirical evidence has been provided by Chen and Zhou (2008). In next section, two new cases will be presented to validate equations (1) to (8), lending further support to the suggestion that hierarchies of cities are identical in cascade structure to network of rivers and size distributions of earthquake energy.

Table 2 Differences between two typical types of hierarchies with cascade structure

| Type | Cascade structure | Interclass relation | Connection | Typical example |
|---|---|---|---|---|
| Real hierarchy | Physical structure | Geometric relation | Concrete connection | River systems |
| Dummy hierarchy | Mathematical structure | Algebraic relation | Abstract connection | Earthquake energy distribution |

## 3. Empirical evidences for urban scaling laws

### 3.1 Cascade structure of USA's hierarchy of cities

The theoretical regularity of city size distributions can be revealed empirically at large scale (Manrubia and Zanette, 1998; Zanette and Manrubia, 1997). The cities in the United States of America (USA) in 2000 are taken as the first example to make an empirical analysis. According to equations (1), (2), and (3), in which the number ratio is taken as $r_n$=2, the 452 US cities with population more than 50,000 can be grouped by population size into 9 levels in the top-down way ($M$=9). The population size is measured by *urbanized area* (UA). The 9 classes compose a hierarchy of cities with cascade structure. The number of cities ($N_m$), the average population size ($P_m$), and the mean urbanized area ($A_m$) in each class are listed in Table 3. The bottom level, namely, the 9th class ($m$=9) is what is called "lame-duck class" by Davis (1978) due to absence of data from the small cities (less than 50,000). Then, the scaling relations between city number and urban population, between city number and urban area, and the allometric relation between urban area and population, can be mathematically expressed with power functions and displayed with double logarithmic plots (Figure 2).

The least squares calculations involved in the data in Table 3 yield a set of mathematical models. The urban size-number scaling relation is

$$\hat{N}_m = 14511580.487 P_m^{-0.974}.$$

The goodness of fit is about $R^2$=0.986, and the fractal dimension is estimated as around $D$=0.974 (Figure 2a). The urban area-number scaling relation is

$$\hat{N}_m = 87304.659 A_m^{-1.213}.$$

The goodness of fit is about $R^2$=0.969, and the fractal parameter is around $d$=1.213 (Figure 2b). The area-population allometric relation is

$$\hat{A}_m = 0.017 P_m^{0.793}.$$

The goodness of fit is around $R^2$=0.993, and the allometric scaling exponent is about $b$=0.793 (Figure 2c). The hat of symbols $N_m$ and $A_m$ (^) denotes the estimated values differing to some extent from the observed values.

The fractal parameters and related scaling exponents can also be estimated by the common ratios. As mentioned above, the number ratio is given *ad hoc* as $r_n$=2. Accordingly, the average size ratio is about $r_p$=2.025, and the average area ratio is around $r_a$=1.768. Thus, consider the formulae given above, $D=\ln r_n/\ln r_p$, $d=\ln r_n/\ln r_a$, $b=\ln r_a/\ln r_p$, we have

$$D \approx \frac{\ln(2)}{\ln(2.025)} \approx 0.983, \quad d \approx \frac{\ln(2)}{\ln(1.768)} \approx 1.217, \quad b \approx \frac{\ln(1.768)}{\ln(2.025)} \approx 0.807.$$

According as the mathematical relationships between different models illuminated in section

2.1, the power-law relations suggest that the hierarchical structure can also be described by a set of exponential functions, i.e., equations (1), (2), and (3). The number law corresponding to equations (1) is known, that is, $N_m=(1/2)e^{\ln(2)m} \approx 0.5e^{0.693m}$. The models of the size law and the area law are in the following forms

$$\hat{P}_m = 41622813.522e^{-0.686m}, \quad \hat{A}_m = 18531.375e^{-0.543m},$$

which correspond to equations (2) and (3). The hat of symbols $P_m$ and $A_m$ (^) indicates the estimated values. The goodness of fit are $R^2=0.991$ and $R^2=0.978$, respectively. The fractal parameters and scaling exponents are estimated as $D \approx 0.693/0.686 \approx 1.010$, $d \approx 0.693/0.543 \approx 1.278$, and $b \approx 0.543/0.686 \approx 0.790$.

**Table 3 The hierarchy of the 452 cities in USA and related measures (2000)**

| Class ($m$) | City number ($N_m$) | Average population size ($P_m$) | Average urban area ($A_m$) |
|---|---|---|---|
| 1 | 1 | 17799861.000 | 8683.200 |
| 2 | 2 | 10048695.500 | 4908.995 |
| 3 | 4 | 4561564.500 | 3923.070 |
| 4 | 8 | 3335242.625 | 2828.796 |
| 5 | 16 | 1690796.250 | 1493.243 |
| 6 | 32 | 815564.656 | 899.782 |
| 7 | 64 | 354537.344 | 451.605 |
| 8 | 128 | 156158.125 | 217.896 |
| 9 | (197) | 69740.228 | 103.053 |

**Source**: The original data come from the US Census Bureau (2002.08.25), only the 452 US cities with population size more than 50,000 are available at: www.demographia.com. **Notes**: (1) The last class of each hierarchy is a lame-duck class. (2) The unit of population is "person", and that of urbanized area is "square kilometers".

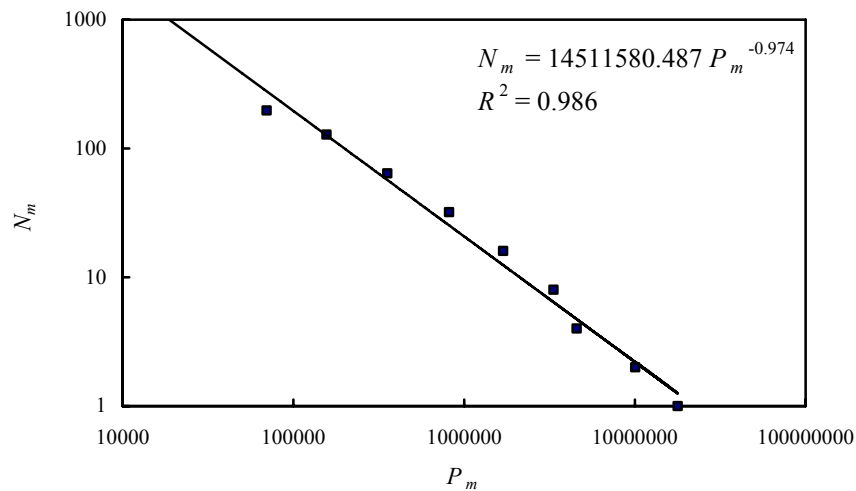

a. The scaling relation between urban population and city number

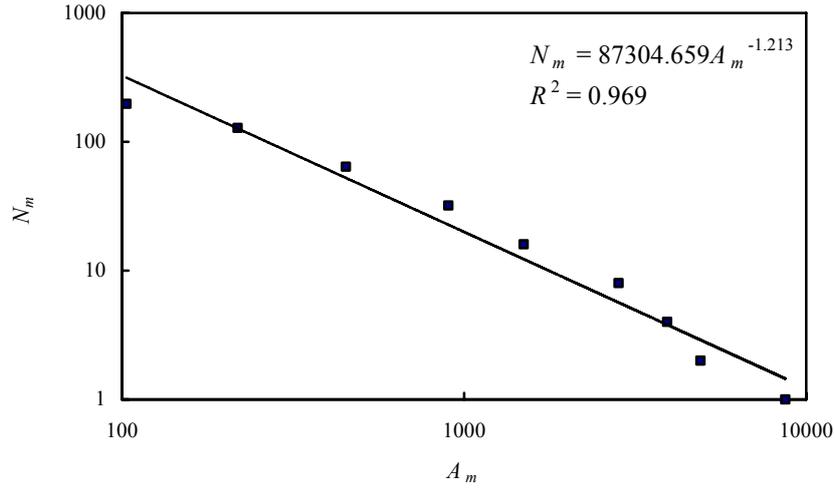

b. The scaling relation between urban area and city number

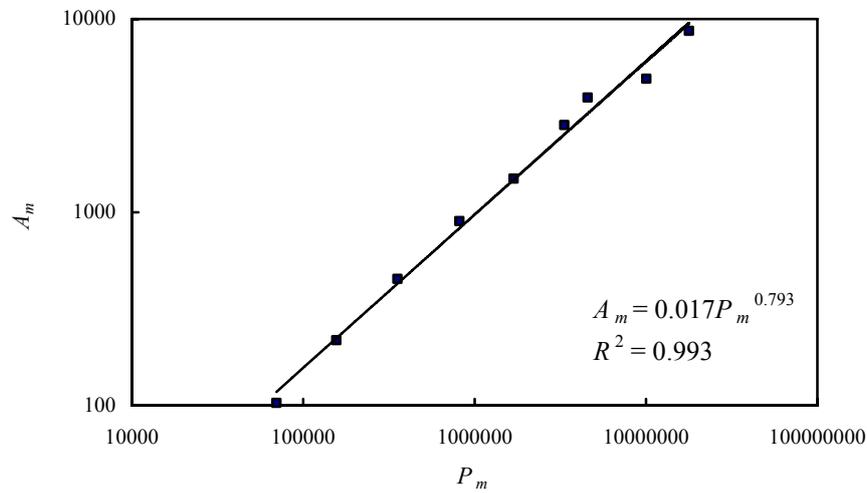

c. The allometric relation between urban area and population

**Figure 2 The scaling patterns for the hierarchy of the 452 cities in America (2000)**

Theoretically, the fractal parameters or scaling exponents of a hierarchy of cities from different ways, including power laws, exponential laws, and common ratios, should be the same as each other. However, in practice, the results based on different approaches are always close to but different from one another due to the uncontrollable factors such as random noises, spatial scale, and degree of system development. The average values of the fractal dimension and allometric scaling exponent can be calculated as $D\approx 0.989$, $d\approx 1.236$, and $b\approx 0.797$.

### 3.2 Cascade structure of PRC's hierarchy of cities

Another large-scale urban system is in the People's Republic of China (PRC). By the similar method, the 660 cities of China in 2005 can be classified by population size into 10 levels ($M$=10). Different from the US cities, the urban area of China's cities is not UA, but the "built-up area (BA)", which is also called "surface area of built district". The city number ($N_m$), the average population size ($P_m$), and the average urban area ($A_m$) in each class are tabulated as follows (Table 4). The bottom level, namely, the 10th class ($m$=10) is also a lame duck class because of undergrowth of small cities. The scaling relations can be expressed with three power functions and are illustrated with log-log plots (Figure 3). For the first two scaling relations, it is better to remove the data point of the lame duck class, which can be regarded as an outlier, from the least square computation in the regression analysis. As is often the case, the power-law relations break down when the scale of observation or systems is too large or too small (Bak, 1996).

Table 4 The hierarchy of the 660 cities in PRC and related measures (2005)

| Class ($m$) | City number ($N_m$) | Average population size ($P_m$) | Average urban area ($A_m$) |
|---|---|---|---|
| 1 | 1 | 1778.420 | 819.880 |
| 2 | 2 | 1182.875 | 956.500 |
| 3 | 4 | 626.830 | 567.405 |
| 4 | 8 | 407.219 | 261.399 |
| 5 | 16 | 237.608 | 183.454 |
| 6 | 32 | 148.627 | 144.776 |
| 7 | 64 | 82.504 | 70.169 |
| 8 | 128 | 43.948 | 44.371 |
| 9 | 256 | 20.544 | 23.189 |
| 10 | (149) | 9.764 | 13.062 |

**Source**: The original data are from *2005 Statistic Annals of China's Urban Construction* published by the Ministry of Housing and Urban-Rural Development of China. **Notes**: (1) The last class of each hierarchy is a lame-duck class. (2) The unit of population is "10 thousands person", and the unit of urban area is "square kilometers".

Analogous to the US case, the least squares computations of the quantities listed in Table 4 give a set of power-law models and exponential models. The urban size-number scaling relation is

$$\hat{N}_m = 14784.254 P_m^{-1.262}.$$

The goodness of fit is $R^2$≈0.995, and the fractal dimension is estimated as $D$≈1.262 (Figure 3a). The urban area-number scaling relation is

$$\hat{N}_m = 28133.543 A_m^{-1.435}.$$

The goodness of fit is $R^2 \approx 0.975$, and the fractal parameter is $d \approx 1.435$ (Figure 3b). The area-population allometric relation is

$$\hat{A}_m = 1.786 P_m^{0.856}.$$

The goodness of fit is $R^2 \approx 0.988$, and the allometric scaling exponent is $b \approx 0.856$ (Figure 3c).

The scaling exponents can also be estimated by number, size, and area ratios. The number ratio is given as $r_n = 2$ (Table 4). Correspondingly, the average size ratio is $r_p \approx 1.796$, and the average area ratio is $r_a \approx 1.638$. In this case, the fractal parameters are estimated as follows

$$D \approx \frac{\ln(2)}{\ln(1.796)} \approx 1.184, \quad d \approx \frac{\ln(2)}{\ln(1.638)} \approx 1.405, \quad b \approx \frac{\ln(1.638)}{\ln(1.796)} \approx 0.842.$$

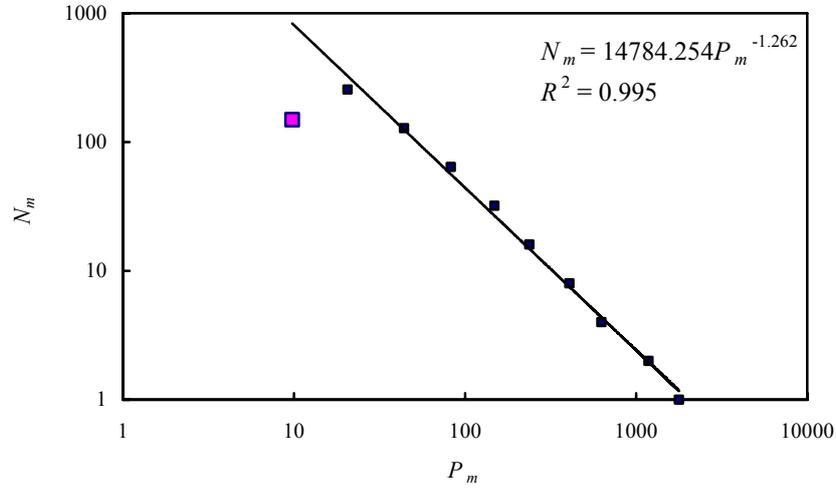

a. The scaling relation between urban population and city number

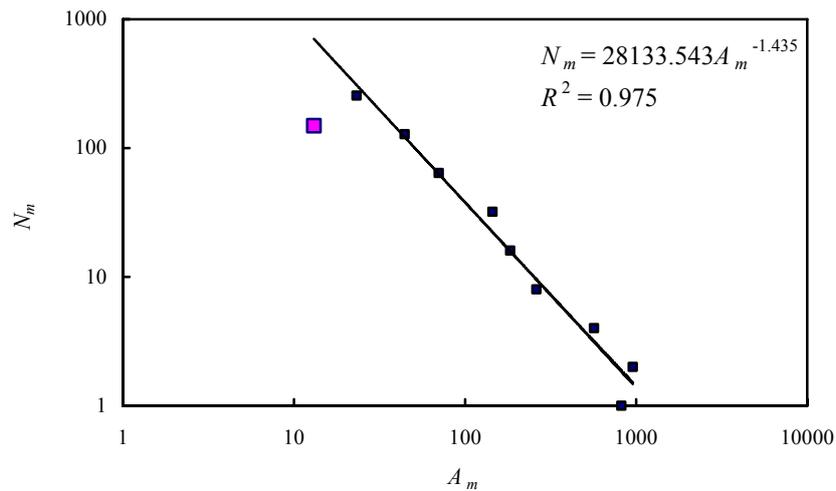

b. The scaling relation between urban area and city number

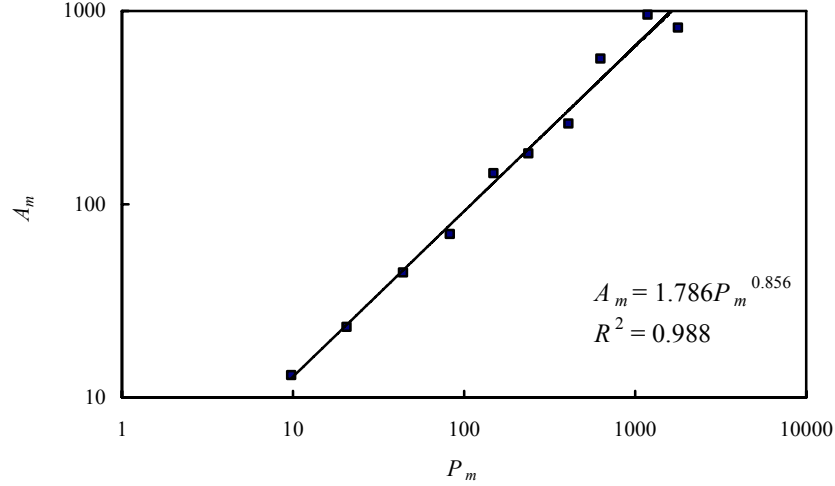

c. The allometric relation between urban area and population

**Figure 3 The scaling patterns for the hierarchy of the 660 cities in China (2005)**

(**Note:** In the first two plots, the data points of the lame duck classes are treated as the outliers, which deviates from the normal scaling range because the small cities in China are of undergrowth).

The above results imply that equations (1), (2), and (3) can also be employed to characterize the hierarchical structure of China's cities. The number law is $N_m=(1/2)e^{\ln(2)m}$. The models of the size and area laws can be expressed as

$$\hat{P}_m = 3726.583e^{-0.568m}, \quad \hat{A}_m = 2030.928e^{-0.486m}.$$

The goodness of fit are $R^2 \approx 0.993$ and $R^2 \approx 0.980$, respectively. The fractal parameters are estimated as $D \approx 0.693/0.568 \approx 1.220$, $d \approx 0.693/0.486 \approx 1.425$, and $b \approx 0.486/0.568 \approx 0.856$. Now, the average values of the fractal parameters or scaling exponents of the hierarchy of the PCR cities from three different ways can be calculated as $D \approx 1.222$, $d \approx 1.422$, and $b \approx 0.851$.

### 3.3 Interpretation of the fractal parameters of urban hierarchies

The fractal property and fractal dimension of a hierarchy of cities can be understood by analogy with the regular fractals such as Cantor set, Koch curve, and Sierpinski carpet (see Appendix 1). A fractal process is a typical hierarchy with cascade structure, and we can model it using the abovementioned exponential functions and power laws, e.g., equations (1) to (6). There are three approaches to estimating the fractal parameters. The first is the regression analysis based on a power law, the second is the least square calculation based on a pair of exponential laws, and the third is numerical estimation based on the common ratios. In theory, the results from these different methods are identical in value to one another. However, for the empirical analysis, they are different to some extent from each other because of the chance factors of urban evolution and local irregularities of hierarchical structure (Table 5). In practice, the method based on the power laws is in common use as it can reflect the scaling relations directly, but the one based on the common ratios is simpler and more convenient. As for the method based on the exponential pair, it can show further information of hierarchical structure. For the random fractals, the more regular the cascade structure of cities is, the more consistent the results from different approaches are. So, in a sense, the degree of consistency of fractal parameter values from the three different methods implies the extent of self-similarity of an urban system.

The fractal dimensions measured by city sizes (population and area) indicate the equality of the city-size distribution. The higher fractal dimension value of an urban hierarchy suggests smaller difference between two immediate classes, while the lower dimension value suggests the larger interclass difference. For the fractal dimension measured by city population $D$, if $r_n>r_p$, then we have $D>1$, otherwise, $D<1$. For the dimension measured by urban area $d$, if $r_n>r_a$, then we have $d>1$, or else, $d<1$. As indicated above, the scaling exponent $b$ is the ratio of $D$ to $d$. it can be understood as an elasticity coefficient. As far as a hierarchy of cities is concerned, the ratio of one dimension to the other dimension (say, $b$) is more important than the value of some kind of fractal dimension (say, $D$ or $d$). If $b>1$, i.e., $D>d$, urban land area grows at a faster rate than that of population (positive allometry), and this suggest that the per capita land area will be more than ever the larger a city becomes; contrarily, if $b<1$, i.e., $D<d$, urban land area grows at a slower rate than that of population (negative allometry), and this implies that the per capita land area will be less the larger a city is. Evidently, if $b=1$, i.e., $D=d$, urban area and population grow at the same rate (isometry), and per capita land area is constant. Thus it can be seen that the scaling exponent can reflect the different types of urban land use: intensive or extensive, saving or wasteful.

Generally speaking, for the cities in the real world, we have $D\leq1$, $d\geq1$. If $D>1$ as given, then $d>D$. Thus $b=D/d\leq1$. Both the USA's cities and PRC's cities satisfy this rule. The similarities and differences between the cities of USA and those of PCR can be found from the parameter values estimated in Table 5. The consistency of fractal parameter values from different approaches is good for the two countries. The fractal dimension value based on city population is less than that based on urban area, i.e., $D<d$. Accordingly, the scaling exponents are less than 1, that is, $b<1$. For the USA's cities, $D≈1$, $d≈1.25$, thus $b≈0.8=4/5$; for the PRC's cities, $D≈1.2$, $d≈1.4$, consequently, $b≈0.857≈6/7$. The different values seem to suggest that the land use of USA's cities is more efficient than that of PRC's cities. However, it should be noted that the differences of parameter values partially results from different measures (say, for urban area, UA differs from BA). Especially, different countries have different definitions about urban area and population size. Anyway, as the whole, the cascade structure of the USA cities is more regular than that of the PRC cities since the $D$ value of USA's cities is closer to 1, and this conforms to Zipf's law.

Table 5 The collected results of the fractals parameters and scaling exponents of the hierarchies of the USA and PCR cities

| Approach | Fractal parameter or scaling exponent | | | | | |
| --- | --- | --- | --- | --- | --- | --- |
| | USA's cities in 2000 | | | PCR's cities in 2005 | | |
| | $D$ | $d$ | $b$ | $D$ | $d$ | $b$ |
| Power law | 0.974 | 1.213 | 0.793 | 1.262 | 1.435 | 0.856 |
| Exponential law | 1.010 | 1.278 | 0.790 | 1.220 | 1.425 | 0.856 |
| Common ratio | 0.983 | 1.217 | 0.807 | 1.184 | 1.405 | 0.842 |
| Mean value | 0.989 | 1.236 | 0.797 | 1.222 | 1.422 | 0.851 |

# 4. Cards shuffling process of urban evolvement

## 4.1 A metaphor of shuffling cards for city distributions

Many evidences show that urban evolution complies with some empirical laws which dominate physical systems. The economic institution, system of political organization, ideology, and history and phase of social development in PCR are different to a great extent from those in USA. However, where the statistical average is concerned, the cities in the two different countries follow the same scaling laws. Of course, the similarity at the large scale admits the differences at the small scale, thus the stability at the macro level can coexist with the variability at the micro level of cities (Batty, 2006). For the self-organized systems, the mathematical models are always based on the macro level, while the model parameters can reflect the information from the micro level. Notwithstanding the difference at the micro level displayed by parameters, the hierarchy of USA cities is the same as that of the PCR cities at the macro level shown by mathematical equations.

All in all, the hierarchy of cities can be described with three exponential models, or four power-law models including Zipf's law. The exponential models reflect the "longitudinal" or "vertical" distribution across different classes, while the power-law models reflect "latitudinal" or "horizontal" relation between two different measurements (say, urban area and population size) (see Appendix 2). The empirical analysis based on both America's and China's cities give support to the argument that, at least at large scale, the hierarchical structure of urban systems satisfy the exponential laws such as equations (1), (2), and (3), or the power laws such as equations (4), (5), and (6). This suggests that the cascade structure of hierarchies of cities can be modeled by the empirical laws which are identical in mathematical form to Horton-Strahler's laws on networks of rivers and Gutenberg-Richter's laws on spatio-temporal patterns of seismic activities.

Urban hierarchy represents the ubiquitous structure frequently observed in physical and social systems. Studies on the cascade structure with fractal properties will be helpful for us to understand how a system is self-organized in the world. In the spatio-temporal evolution of cities in a region, there are at least two kinds of the unity of opposites. One is the global target and local action, and the other is determinate rule (at the macro level) and the random behavior (at the micro level). To interpret the mechanism of urban evolution and the emergence of rank-size patterns, a deck shuffling theory is proposed here. A regional system (a global area) consists of many subsystems (local areas), and each subsystem can be represented by a card. The card shuffling process symbolizes the introduction of randomicity or chance factors into evolution of regions and cities. The model of shuffling cards is only a metaphor, and the logic between this model and real systems of cities is not very significant.

Suppose there are many blank cards. We can play a simple "game" step by step as follows (Figure 4).

**Step 1, put these blank cards in "apple-pie" order to form a rectangle array.** For simplicity, let the number of cards in the array be $u \times v$, where $u$ and $v$ are positive integers. There is no interspace or overlap between any two cards (Figure 4a). As a sketch map, let's take $u=v=3$ for instance.

**Step 2, fix these ordered blanks cards for the time being.** Then draw a hierarchy of "cities" to form a regular network with cascade structure in light of equations (1), (2), and (3). Let the size distribution of cities follow Zipf's law with $q=1$ (Figure 4b). In this instance, both the mathematical structure and physical structure can be described with the exponential laws or power laws given above.

**Step 3, shuffle cards. Note that these cards are not blank and form a deck now.** Unfix and mix these cards together, then riffle these cards again and again at your pleasure (Figure 4c). Finally, the cards are all jumbled up so that the spatial order disappears completely.

**Step 4, rearrange the cards closely.** Take out cards at random one by one from the deck, and place them one by one to form a $u \times v$ array again (Figure 4d). The result is very similar to the map of real cities.

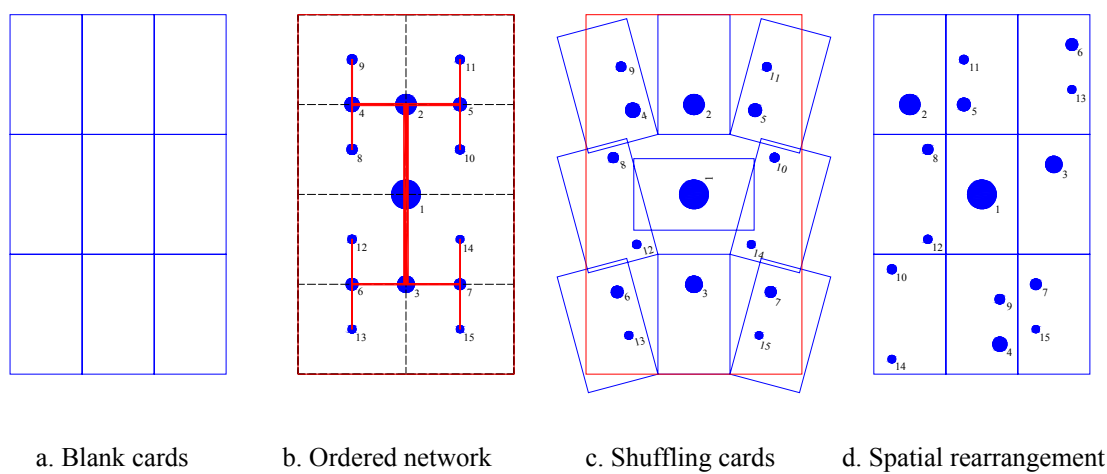

a. Blank cards     b. Ordered network     c. Shuffling cards     d. Spatial rearrangement

**Figure 4 A sketch map of shuffling cards of network of cities**

(**Note:** The sizes of cities conform to the rank-size rule, equation (9). The numbers denote the rank of cities. The network in Figure 4b is constructed according to the $2^n$ principle, but only the first four classes are shown here).

Examining these shuffled cards in array, you will find no ordered network structure of "cities" anymore. The physical structure of the network of "cities" may not follow the exponential laws and power laws yet. To reveal the hidden order, we must reconstruct the hierarchy according to certain scaling rule. Thus the physical cascade structure changes to the mathematical cascade structure, and then the regular physical hierarchy can be replaced by the dummy hierarchy (Table 6). The central place models presented by Christaller (1933/1966) represent the regular hierarchy, while the real cities in a region, say, America or China, can be modeled by a dummy hierarchy. In particular, in step 2, the cities are arranged by the ideas of recursive subdivision of space and cascade structure of network (Goodchild and Mark, 1987). The spatial disaggregation and network development can be illustrated by Figure 5 (Batty and Longley, 1994). After shuffling "cards", the regular geometric pattern of network structure is destroyed, but the mathematical pattern is preserved and can be disclosed by statistical average analysis at large scale.

**Table 6 Comparison of hierarchy model between the cases before and after shuffling cards**

| Item | Before shuffling cards | After shuffling cards |
| --- | --- | --- |
| Mathematical cascade structure | Exist | Keep |
| Physical cascade structure | Exist | Fade away |
| Fractal property | Regular fractal | Random fractal |
| Zipf distribution | Exist | Keep |
| Network type | Real hierarchy | Dummy hierarchy |

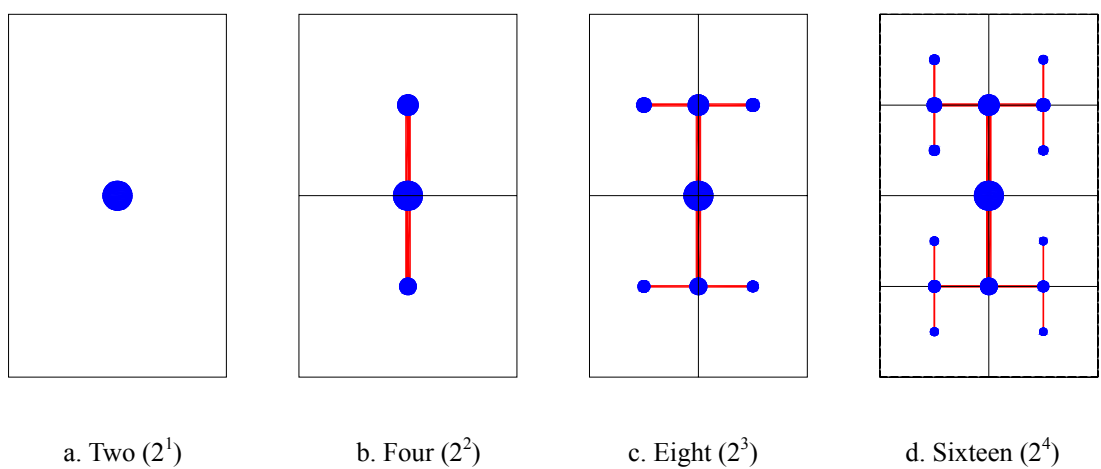

a. Two ($2^1$)   b. Four ($2^2$)   c. Eight ($2^3$)   d. Sixteen ($2^4$)

**Figure 5 Spatial disaggregation and network growth (the first four steps) (by referring to Batty and Longley, 1994)**

## 4.2 Zipf's law as a signature of hierarchical structure

After shuffling "cards", the regularity of network structure will be lost, but the rank-size pattern will keep and never fade away. In this sense, Zipf's law is in fact a signature of hierarchical structure. This can be substantiated by the empirical cases. For the 482 US cities with population over 50,000, a least square calculation yield such a model

$$\hat{P}(\rho) = 52516701.468 \rho^{-1.125}.$$

The goodness of fit is about $R^2$=0.989, and the fractal dimension of urban hierarchy is estimated as around $D=1/q \approx 1/1.125 \approx 0.889$. Since the scaling relation of size distributions often break down when the scale is too large or too small (Batty, 2008; Chen and Zhou, 2008), we should investigate the scaling range between certain limits of sizes. For the 594 PRC cities with population size over 100,000, which approximately form a line on log-log plot (Figure 6), the rank-size model is

$$\hat{P}(\rho) = 48416658.931\rho^{-0.925}.$$

The goodness of fit is $R^2 \approx 0.979$, and the fractal dimension is estimated as about $D=1/q\approx 1/0.925 \approx 1.081$. Please note that the sample size for the rank-size analysis here differs to some extent from that for the hierarchical analysis in section 3.2. Despite some errors of parameter estimation, the mathematical structure of urban hierarchy is indeed consistent with the Zipf distribution.

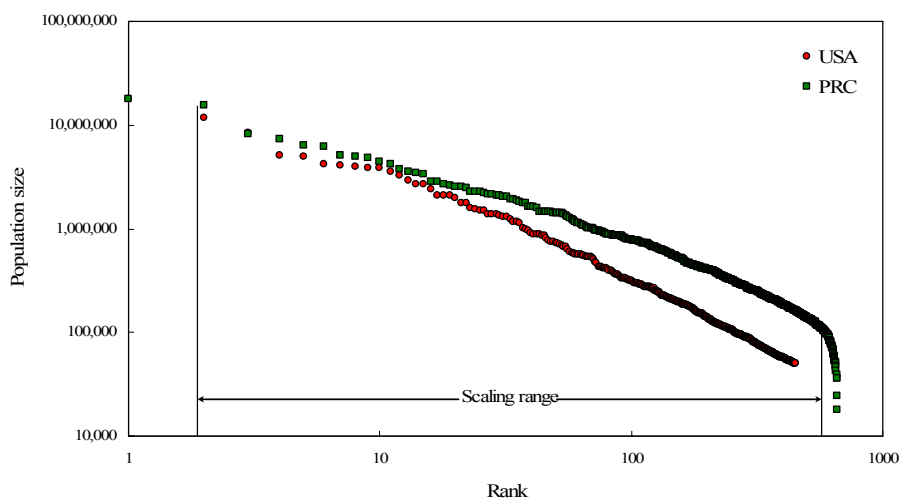

**Figure 6 The rank-size patterns of US cities in 2000 and PRC cities in 2005**

## 4.3 Symmetry breaking and reconstruction of urban evolution

The idea from shuffling cards can be employed to interpret urban phenomena such as the relationship between central place models and spatial distribution of human settlements in the real world. The central place models suggest the ideal hierarchies of human settlements with cascade structure (Christaller, 1933/1966), while the spatial patterns of real cities and towns is of irregularity and randomicity. If the actual systems of cities are as perfect as the models of central places, they will yield no new information for human evolution. Urban systems can be regarded as the consequences of the standard central place systems after "shuffling cards". When the cards with central place patterns are shuffled, the ordered network patterns are thrown into confusion, but the rank-size pattern never changes. To reveal the regularity from urban patterns with irregularity, we have to model hierarchy of cities and then construct a dummy network (Figure 7).

The process of shuffling cards is a metaphor of symmetry breaking of *apriori* ordered network.

Owing to symmetry breaking, chance factors are introduced into the determinate systems, thus randomicity or uncertainty comes forth (Prigogine and Stengers, 1984). In a sense, it is symmetry breaking that lead to complexity. Precisely because of this, we have illimitable information and innovation from complex systems. The question is how to disclose the simple rules behind the complex behaviors of complex physical and social systems. A possible way out is to reconstruct symmetry by modeling hierarchies (Figure 7).

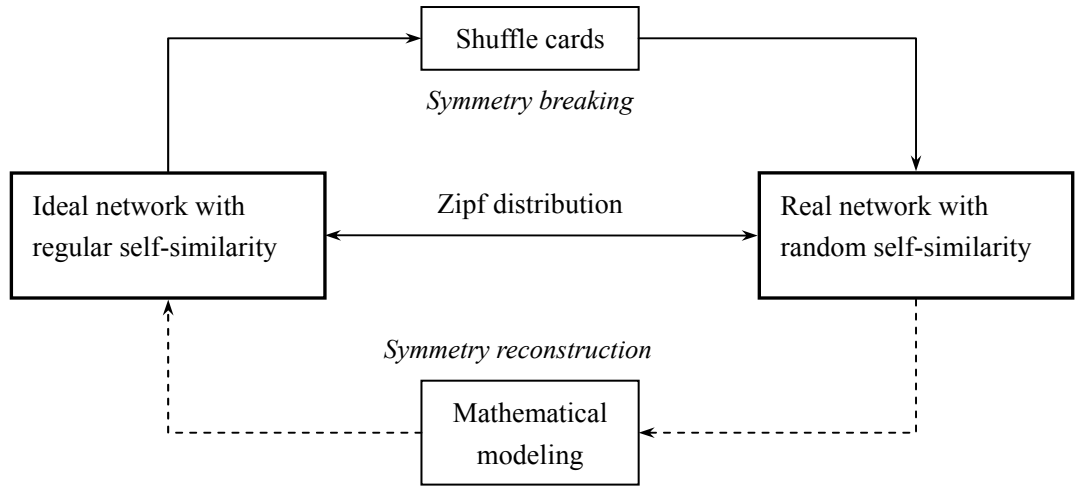

**Figure 7 A schematic diagram of symmetry breaking and reconstruction of network of cities**

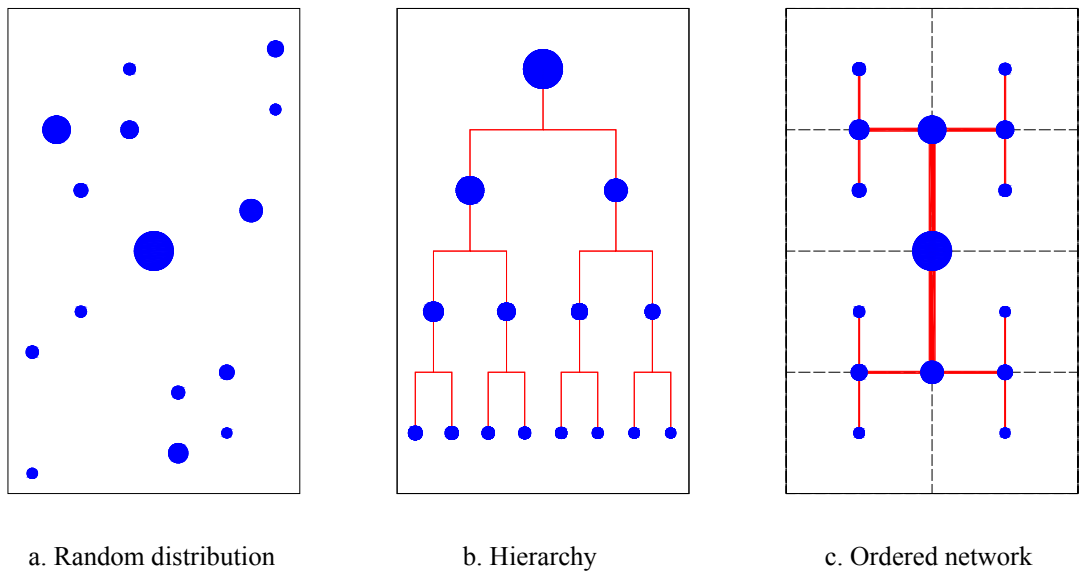

a. Random distribution  b. Hierarchy  c. Ordered network

**Figure 8 Hierarchical structure as a knowledge link between the apriori ordered network and the empirical random distribution of cities (the first four classes)**

A hierarch with cascade structure can be treated as a "mathematical transform" from real cities to the regular cities (Figure 8). Suppose there is a random pattern reflecting the spatial distribution of cities (Figure 8a). This pattern represents the systems of cities after "shuffling cards" (Figure 4d). The city size distribution of this system follows Zipf's law. Let the number ratio $r_n$=2. Then we can construct a hierarchy with cascade structure (Figure 8b). This hierarchy is in fact a dummy network of cities. By the principle of recursive subdivision of geographical space (Batty and Longley, 1994; Goodchild and Mark, 1987), we can reconstruct an ordered network of cities (Figure 8a). This model on systems of cities can represent the regular network before "shuffling cards" in the *apriori* world (Figure 4b, Figure 5d).

## 5. Discussion and conclusions

Zipf's law used to be considered to contradict the hierarchy with cascade structure. Many people think that the inverse power law implies a continuous distribution, while the hierarchical structure seems to suggest a discontinuous distribution. In urban geography, the rank-size distribution of cities takes on a continuous frequency curve, which is not consistent with the hierarchical step-like frequency distribution of cities predicted by central-place theory (Christaller, 1933/1966). However, the problem of the contradiction between the Zipf distribution and the hierarchies of central places has been resolved by different theories and methods (e.g. Allen, 1997; Chen and Zhou, 2003; Chen and Zhou, 2004; Prigogine and Stengers, 1984). In fact, the size distributions of urban places in the real world always appear as approximately unbroken frequency curves rather than the stair-like curves. The step-like hierarchical structure of is based on spatial symmetry, but according to dissipative structure theory, such a regular hierarchical distribution as central place patterns is very infrequent in actual case because that the spatial symmetry is always disrupted by the historical, political, and geographical factors (Prigogine and Stengers, 1984). What is more, the regular hierarchical structure is not allowed by the nonlinear dynamics of urbanization (Chen and Zhou, 2003), and the simple fractal structure of urban hierarchies is often replaced by the multifractal structure (Chen and Zhou, 2004). The multifractals of urban hierarchies suggest an asymmetrical hierarchy of cities, which differs from the hierarchical systems in central place theory.

Therefore, the hierarchical models are for ever based on the idea of statistical average rather than reality or observations. In terms of statistical average, the rank-size distribution can always be transformed into a hierarchy with cascade structure. However, the traditional hierarchical structure predicted by central place theory cannot be transformed into the rank-size distribution. On the other hand, the size distributions in the real world support Zipf's law and the hierarchical model based on statistical average instead of the step-like hierarchical distribution. Consequently, a conclusion can be drawn that the absolute hierarchy should be substituted by the statistical hierarchy associated with the rank-size distribution. Precisely based on this concept, the metaphor of shuffling cards is proposed to interpret the urban evolution coming between chaos and order.

To sum up, Zipf's law is a simple rule reflecting the ubiquitous general empirical observations in both physical and human fields. However, the underlying rationale of the Zipf distribution remains to be revealed. This paper tries to develop a model to illuminate the theoretical essence of the rank-size distribution: the invariable patterns of evolutive network or hierarchy. The hierarchy with cascade structure provides us with a new way of looking at the rank-size distribution. The hierarchy can be characterized by both exponential laws and power laws from two different perspectives. The exponential models (e.g. the generalized $2^n$ rule) and power-law models (e.g. the rank-size rule) of cities represent the general empirical laws. Studies on the human systems of cities will be instructive for us to understand physical phenomena such as rivers and earthquakes. By analogy with cities, we can understand river networks and earthquake behaviors and all the similar physical and social systems with hierarchical structure from new perspectives.

The theory of shuffling cards is not an underlying rationale, or an ultimate principle. As indicated above, it is a useful metaphor. The idea from cards shuffling is revelatory for us to find new windows, through which we can research the mechanism of the unity of opposites such as chaos and order, randomicity and certainty, and complexity and simplicity. A conjecture or hypothesis is that complex physical and social systems are organized by the principle of dualistic structure. One is the mathematical structure with regularity, and the other is the physical structure with irregularity or randomicity. The mathematical structure represents the *apriori* structure before shuffling cards, while the physical structure indicates the empirical structure after shuffling cards. A real self-organized system always tries to evolve from the physical structure to the mathematical structure for the purpose of optimization. In short, in the process of "shuffling cards" of urban system, there is an invariable and invisible pattern. That is the rank-size distribution dominated by Zipf's law. To bring to light the latent structure and basic rules of urban evolution, further studies should be made on the rank-size pattern through proper approach in the future.

**Acknowledgements**: This research was sponsored by the National Natural Science Foundation of China (Grant No. 40771061). The support is gratefully acknowledged.

# Appendices

## A1 Fractals, hierarchies, and fractal dimension

A regular fractal is a typical hierarchy with cascade structure. Let's take the well-known Cantor set as an example to show how to describe this hierarchical structure and how to calculate its fractal dimension (Figure A1). We can use two measurements, the length ($L$) and number ($N$) of fractal copies in the $m$th class, to characterize the fractal hierarchy. Thus, we have two exponential functions such as

$$N_m = N_1 r_n^{m-1} = \frac{N_1}{r_n} e^{(\ln r_n)m} = N_0 e^{\omega m}, \tag{A1}$$

$$L_m = L_1 r_l^{1-m} = L_1 r_l e^{-(\ln r_l)m} = L_0 e^{-\psi m}, \tag{A2}$$

where $m$ denotes the ordinal number of class ($m=1, 2, \ldots$), $N_m$ is the number of the fractal copies of a given length, $L_m$ is the length of the fractal copies in the $m$th class, $N_1$ and $L_1$ are the number and length of the initiator ($N_1=1$), respectively, $r_n$ and $r_l$ are the **number ratio** and **length ratio** of fractal copies, $N_0=N_1/r_n$, $L_0=L_1 r_l$, $\omega=\ln(r_n)$, $\psi=\ln(r_l)$. Apparently, the common ratios are

$$r_n = \frac{N_{m+1}}{N_m} = \frac{N_1 r_n^m}{N_1 r_n^{m-1}} = 2, \quad r_l = \frac{L_m}{L_{m+1}} = \frac{L_1 r_l^{1-m}}{L_1 r_l^{-m}} = 3.$$

From equations (A1) and (A2), we can derive a power law in the form

$$N_m = k L_m^{-D}, \tag{A3}$$

in which $k=N_1 L_1^D$ is the proportionality coefficient, and $D$ is the fractal dimension of the Cantor set. Based on the power law, the fractal dimension can be expressed as

$$D = -\frac{\ln(N_{m+1}/N_m)}{\ln(L_{m+1}/L_m)}, \tag{A4}$$

Based on the exponential models, the fractal dimension is

$$D = \frac{\omega}{\psi}. \tag{A5}$$

Based on the common ratios, the fractal dimension is

$$D = \frac{\ln r_n}{\ln r_l}. \tag{A6}$$

In light of both the mathematical derivation and the empirical analysis, we have

$$D = -\frac{\ln(N_{m+1}/N_m)}{\ln(L_{m+1}/L_m)} = \frac{\ln r_n}{\ln r_l} = \frac{\omega}{\psi} = \frac{\ln(2)}{\ln(3)} \approx 0.6309.$$

This suggest that, for the regular fractal hierarchy, the fractal dimension can be computed by using exponential functions, power function, or common ratios, and all these results are equal to one another.

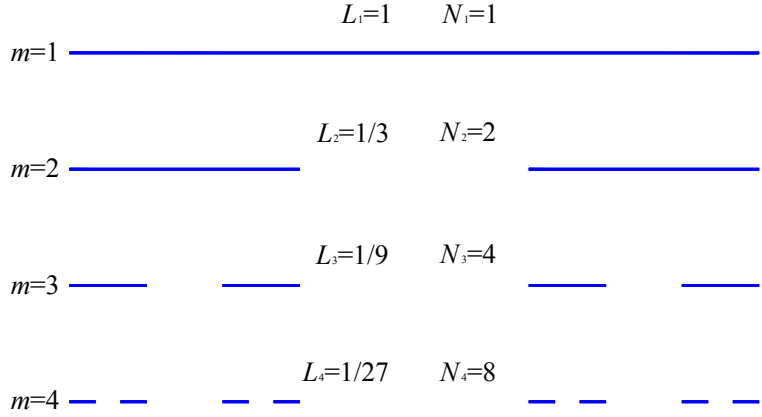

**Figure A1 The Cantor set as a hierarchy with cascade structure (the first four classes)**

Further, if we introduce the third measurement, the "weight" of the fractal copies ($W$), to characterize the Cantor set, we have

$$W_m = W_1 r_w^{1-m} = L_1 r_w e^{-(\ln r_w)m} = W_0 e^{-\upsilon m}, \tag{A7}$$

where $W_m$ is the weight of fractal copies in the $m$th class, $W_1$ is the weight of the initiator ($W_1=1$), $r_w=W_m/W_{m+1}$ is the *weigth ratio* of fractal copies, $W_0=W_1 r_w$, $\upsilon=\ln(r_w)$. In this instance, we can derive an allometric scaling relation such as

$$W_m = \eta L_m^b, \tag{A8}$$

where $\eta=W_1 L_1^{-b}$ is the constant coefficient, and $b$ is the allometric scaling exponent, which can be defined as

$$b = \frac{\ln(W_m/W_{m+1})}{\ln(L_m/L_{m+1})} = \frac{\ln r_w}{\ln r_l} = \frac{\upsilon}{\psi} = \frac{\ln(3)}{\ln(3)} = 1.$$

This implies that, for the Cantor set, the measurement of weight is in fact equivalent to the length, and thus is unnecessary for this fractal body.

The mathematical description and fractal calculation of the Cantor set can be generalized to other regular fractals such as Koch snowflake and Sierpinski gasket or even to the route from bifurcation to chaos. As a simple fractal, the Cantor set fails to follow Zipf's law. However, if we substitute the multifractal structure for monofractal structure, the multi-scaling Cantor set will empirically follow Zipf's law.

## A2 Longitudinal relations and latitudinal relations of hierarchies

The longitudinal relations are the associations across different classes, while the latitudinal relations are the correspondences between different measures such as city population size and urban area. These relations can be illustrated with the following figure (Figure A2).

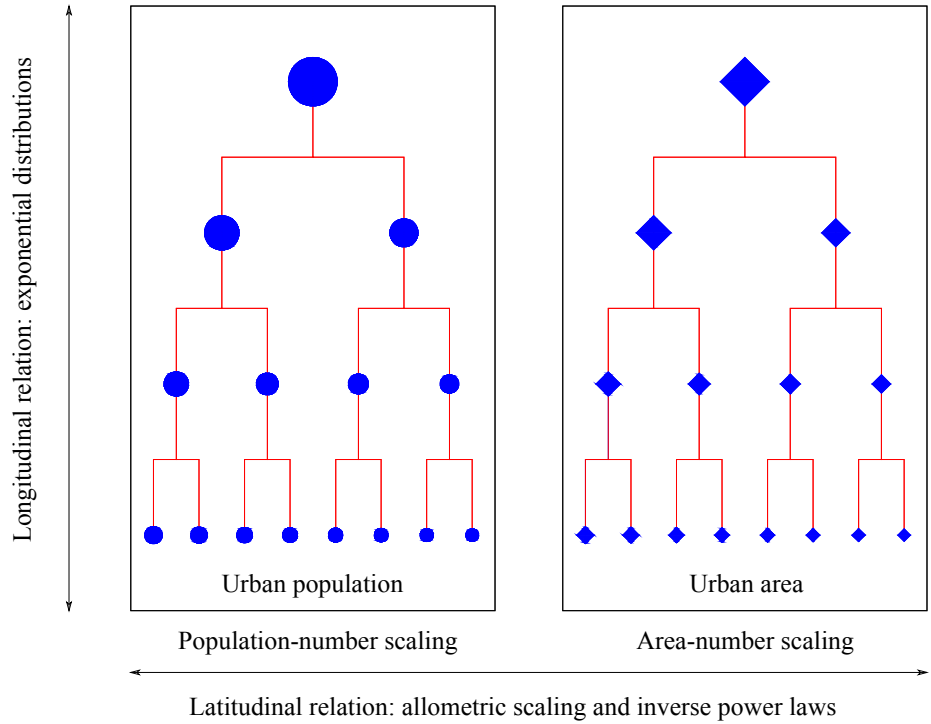

**Figure A2 A schematic diagram on the longitudinal relations and latitudinal relations of urban hierarchy (the first four classes)**